\documentclass{emulateapj}

\newcommand{\kms}{\,km\,s$^{-1}$}      
\newcommand{\tm}{\tablenotemark} \newcommand{\tn}{\tablenotetext}
\newcommand{\fuse}{\emph{FUSE}}

\begin{document}
\shorttitle{Measuring Noisy Spectra}
\shortauthors{Fox, Savage, \& Wakker}

\title{Measurement of Noisy Absorption Lines using the Apparent
  Optical Depth Technique}
\author{Andrew J. Fox, Blair D. Savage, \& Bart P. Wakker,}
\affil{Department of Astronomy, University of Wisconsin -
Madison, 475 North Charter St., Madison, WI 53706}
\email{fox@astro.wisc.edu}

\begin{abstract}
To measure the column densities of interstellar and intergalactic gas
clouds using absorption line spectroscopy, the apparent
optical depth technique (AOD) of \citet{SS91} can be used instead of a
curve-of-growth analysis or profile fit. We show that the AOD
technique, whilst an excellent tool when applied to data with good
S/N, will likely overestimate the true column densities when
applied to data with low S/N. This overestimation
results from the non-linear relationship between the flux falling on a
given detector pixel and the apparent optical depth in that pixel. 
We use Monte Carlo techniques to investigate the amplitude of this
overestimation when working with 
data from the {\it Far Ultraviolet Spectroscopic Explorer} (\fuse)
and the Space Telescope Imaging Spectrograph (STIS), for a
range of values of S/N, line depth, line width, and rebinning.
AOD measurements of optimally sampled, resolved lines are accurate to
within 10\% for {\fuse}/LiF and STIS/E140M data with S/N$\gtrsim$7 per
resolution element. 
\end{abstract}

\section{Introduction}
The apparent optical depth (AOD) method \citep{SS91, SS92} has gained
substantial popularity as a means of converting velocity-resolved flux
profiles into column density measurements for interstellar and
intergalactic absorption lines. The method offers a quick 
and convenient way of determining reliable interstellar column
densities from unsaturated lines without having to follow a full
curve-of-growth  
analysis or detailed component fit, and without demanding prior knowledge
of the component structure. Accurate column densities are important
for studies of elemental abundances and physical conditions in diffuse
gas in space.

In the AOD method, a velocity-resolved flux profile $F(v)$ is
converted to an apparent optical depth profile $\tau_a(v)$ using the relation:
\begin{equation}
\tau_a(v)={\mathrm ln}[F_c(v)/F(v)],
\end{equation}
where and $F(v)$ and $F_c(v)$ are the observed line and continuum fluxes at
velocity $v$, respectively. This apparent optical depth profile can
then be converted to an apparent column density profile according to
\begin{equation}
N_a(v)=3.768\times10^{14}(f\lambda)^{-1}\tau_a(v)\,
\mathrm{cm^{-2}\,(km\,s^{-1})^{-1}},
\end{equation}
where $f$ is the oscillator strength of the transition and
$\lambda$ is the transition wavelength in Angstroms
\citep{SS91}. The total apparent column density between two velocity
limits $v-$ and $v+$ is then simply $N_a=\int_{v-}^{v+}
N_a(v)\mathrm{d}v$. When two lines of different strength of the same
ionic species are 
available, the AOD method can also be used to assess and correct for
the level of saturation 
in the data, by comparing the apparent column density derived
from the stronger line with that derived from the weaker line
\citep{SS91, Je96}.

The AOD method was originally developed for application to
measurements of reasonably high signal-to-noise spectra (S/N$\gtrsim$20
per resolution element). The technique offers excellent results in
these cases. However, spectroscopists are now using the
method even when analyzing relatively low S/N data \citep[e.g.][]{Wa03}. This
paper deals with accounting for a systematic error that is introduced
when applying the AOD method to noisy data, which can lead to an
overestimation of the true column density of the absorbing 
species. The error arises because the logarithmic
relationship between $\tau_a(v)$ and $F(v)$  (Eq. 1) will distort a
Poisson noise component in the flux when converting into column
density space, giving undue weight to the pixels where the noise has
resulted in high optical depths; this effect tends to exaggerate the
estimate of $N_a$. 
We present simulations investigating the accuracy of $N_a$
determinations as a function of
line depth and S/N, together with the separate effects of
line width and rebinning. 

\section{Simulations}
We ran Monte Carlo simulations of the measurement of noisy
absorption lines to investigate the amount by which noise in the data leads
to overestimation of $N_a$.
Our models were run with two simulated experimental setups, representing
the {\it Far Ultraviolet Spectroscopic Explorer} (\fuse) LiF
\citep{Sa00} and Space Telescope Imaging Spectrograph (STIS) E140M
\citep{Wo98} configurations.
The \fuse/LiF setup is 
parameterized by a Gaussian instrumental line spread function with FWHM=20\kms\
($\sigma_{ins}=8.5$\kms) and velocity pixels 2.0\kms\ wide. The
STIS/E140M configuration has FWHM=6.8\kms\ ($\sigma_{ins}=2.9$\kms) and
3.2\kms\ pixels. Our models simulate the measurement of 
an absorption line whose intrinsic optical depth
profile is a single-component Gaussian, centered for
convenience at 0\kms, and normalized to a peak optical depth of $\tau_0$,
i.e. $\tau(v)=\tau_0e^{-v^2/2\sigma_{line}^2}$. We run models for
two line width 
cases: resolved ($\sigma_{line}=2\sigma_{ins}$) and marginally resolved
($\sigma_{line}=\sigma_{ins}$)
The AOD method is known to
underestimate the column density for unresolved lines, when no
allowance is made for the effects of unresolved saturation [see Table
5 of \citet{SS91}].

For a grid of different intrinsic central line depths $d_{line}$
we converted the optical depth profile to a normalized flux profile using 
$F_{norm}=e^{-\tau(v)}$, where the
intrinsic central optical depth $\tau_0=-\mathrm{ln}(1-d_{line})$.
The flux profiles were then convolved
with the instrumental broadening function; the effect of this process
is to decrease the intrinsic line depth to an apparent line depth ($d$), and
increase the intrinsic line width to an apparent line width ($\sigma$).
We then added different levels of random noise (drawn from a Gaussian
distibution) to the data, scaled so
that the S/N in the continuum reached the desired levels when rebinned
to pixels equal in size to the resolution element. In the line, we
scaled down the random noise contribution by $\sqrt F$, in accordance with
Poisson statistics. 

We then rebinned the data by different 
factors, converted the profiles to $N_a(v)$ space, and finally
integrated the profiles over velocity to yield $N_a$, as a function of
line depth, noise, and rebinning factor. The velocity integration limits used
are $\pm2.5\sigma$ around the line center, corresponding to the points
where the line recovers to 96\% of the continuum in the absence of
noise. These limits were chosen so as to reproduce the velocity
integration ranges used in practice, determined by
comparing the line widths and velocity integration ranges from a large
spectroscopic data set 
\citep[the \fuse\ \ion{O}{6} survey;][]{Wa03}. For each model run, we
determined the ratio of $N_a$ to $N_{true}$, where  
$N_{true}$ is the true column density of the simulated component, then repeated
the process 500 times; the mean overestimation factor
$<N_a>/N_{true}$, together with its dispersion, is then used
in our results.

For very optically thick lines, the random error can take the flux
below zero, corresponding to negative $N_a(v)$, which has no physical
meaning. Therefore, a
cutoff must be used, whereby all points with $F(v)$ less than some
value are placed at that value (in our case, 1\% of the
continuum). The cutoff value selected will depend on the reliability
of the scattered light and background corrections for the particular
instrument providing the observations. 

\section{Results}
Figure 1 shows a graphical example of the overestimation process for
a resolved line with an intrinsic depth of 0.8, measured with \fuse.
In the left column we show how the spectra would appear at different S/N
levels, ranging from S/N=$\infty$ at the bottom to S/N=5 at the
top. Note that we quote the S/N per resolution element, rather than
the S/N at the level of rebinning
shown. The two are related by (S/N)$_{res. elem.}$=$\sqrt{}$(number of bins per
resolution element) $\times$(S/N)$_{bin}$. We verified this relation
by measuring the S/N of the model spectra at different
binning levels. In the next three columns we show the profiles in
apparent column density space after different levels of rebinning. 

In Figure 2 we display the results of different parameter  
cases using a color-coded plot: resolved
lines with \fuse\ (top left), marginally resolved lines with \fuse\
(top right), 
resolved lines with STIS (bottom left), and marginally resolved lines with STIS
(bottom right). Within each panel, we show the overestimation factor 
as a function of signal-to-noise ratio and apparent line depth (both
easily measurable quantities), for three different rebinning
cases. In each panel, we have not shaded in regions 
that correspond to non-significant detections, i.e. those with
$W_{\lambda}<3\sigma(W_{\lambda})$; the minimum depth line that can be
considered significant is proportional to (S/N)$^{-1}$, since at low
S/N shallow lines will not be distinguished from the noise.
Using Figure 2 an observer can find the likely error on $N_a$ in a
line with a given depth, width, and S/N, and correct for it. 
The overestimation factors for a range of cases are also presented in
Tables 1 and 2, for {\it FUSE}/LiF and STIS/E140M data,
respectively. Upon request we can also provide an IDL code that 
returns the overestimation factor when given the measured S/N,
apparent line depth, and apparent line width.

\section{Discussion}
The effects of several parameters combine to produce the trends seen
in Tables 1 and 2 
and Figures 1 and 2: S/N, level of rebinning, spectral resolution,
line width, and line depth. We discuss each separately in this section.

\begin{itemize}

\item {\bf Low S/N and Rebinning} -- the non-linear distortion of the
  $N_a(v)$ profile and resulting overestimate of $N_a$ becomes worse as the
  S/N decreases (Fig. 1). Rebinning the spectrum before measurement
  can lessen the distortion, since the rebinning process increases the
  effective S/N. Once the data is rebinned to a level coarser than the 
  resolution element, the gain in the $N_a$ accuracy is made at the
  expense of resolution, so that kinematic information is given
  up. However, if the absorption line before rebinning is fully or
  marginally resolved, and if one is more interested in obtaining an accurate
  column density than obtaining knowledge about the detailed shape of
  the line profile, then overbinning is justified at low S/N. AOD
  measurements made at optimum sampling (two rebinned pixels per
  resolution element) on {\it FUSE} and STIS data are overestimated by less
  than 10\% for S/N per resolution element $\gtrsim$7. 
  For data with lower S/N, rebinning beyond
  the resolution element will likely increase the accuracy of the
  derived column density. 
  
\vspace{-0.3cm}
\item {\bf Lack of Resolution and Line Width} -- for marginally
  resolved lines, the apparent column density  
  tends to underestimate the true column density at high line depths (purple
  shading on the right hand panels in Figure 2), since the
  high optical depth points are smoothed out to lower
  apparent optical depths by the instrumental blurring. This effect
  therefore works in the opposite sense to the S/N effect. The more
  unresolved the line is (i.e., the lower $\sigma_{line}/\sigma_{ins}$), the
  worse this effect becomes. Even full saturation can be hidden by the
  effects of instrumental broadening. \citet{SS91} and \citet{Je96}
  discuss how this unresolved saturation can be corrected for when
  observations exist for more than one line of an ion.
  For resolved lines, increasing the ratio $\sigma_{line}/\sigma_{ins}$ above
  2 increases the number of pixels susceptible to noise and hence
  increases $<\!N_a\!>/N_{true}$ slightly for a given S/N and line
  depth (by a few percent if
  $\sigma_{line}/\sigma_{ins}=4$); however, since the dependency on
  width is weak, we present results for two line width cases only.
 
\vspace{-0.3cm}
\item {\bf Line Depth} -- for a given S/N and line width, the
  overestimation factor is worse for shallow lines than for deep
  lines. This is because the noise is Poissonian, so a pixel
  with low optical depth (shallow) has a higher error on the flux than a pixel
  with high optical depth (deep). For very
  deep lines, the cutoff used to eliminate negative pixels
  {\it underestimates} the optical depth in very opaque parts of 
  the profile. This effect works in the opposite sense to the S/N
  effect, so that measurements of very deep ($F(v)/F_c(v)\le0.10$)
  lines at low S/N 
  are distorted by two competing effects. Note that if the noise were
  random rather than Poissonian (i.e. independent of the flux), the
  overestimation effect on deep lines would become worse 
  since the flux errors deep in the lines are larger.
 
\end{itemize}

In Figure 3 we show the
histogram of $N_a/N_{true}$ for one particular run (\fuse/LiF,
$\sigma_{line}=2\sigma_{ins}$, S/N=8, intrinsic depth=0.5, 500 runs)
for three levels 
of rebinning. The overestimation factor has a 
Gaussian distribution, with a mean 
that decreases with rebinning. For any given absorption profile, the
overestimation can take a range of values. We therefore advise a
conservative error estimate when correcting for the overestimation.

We note that the
non-linear S/N distortion effect discussed here can influence the
detailed intercomparisons of $N_a(v)$ profiles of the same ion.
Therefore, corrections for line saturation based on comparing
$N_a(v)$ profiles of weak and strong lines may not be valid for low
S/N observations.

In their survey of \ion{O}{6} absorption in the Galactic halo and
high-velocity clouds, \citet{Wa03} rebinned \fuse\ data with S/N$<$10 per
resolution element by 5 pixels, and noisier data by up to 20
pixels, before measurement with the AOD technique. Our modelling shows
that this was the correct procedure for measuring accurate \ion{O}{6}
column densities, and we endorse a similar treatment of low S/N data
in future studies. Without this rebinning, the  
column density measurements along sight lines with noisy spectra
could be inflated by more than 20\%.

{\bf Acknowledgements} The authors wish to thank Ken Sembach and Chris Howk for
useful discussions. This research has been supported by NASA through
grant NNG04GC70G to BDS and grant NAG5-7444. BDS also acknowledges
support from the University of Wisconsin Graduate School.

\begin{deluxetable}{lcccc ccc}
\tablewidth{0pt}
\tabcolsep=4pt
\tabletypesize{\scriptsize}
\tablecaption{Simulation Results - \fuse/LiF}
\tablehead{\multicolumn{2}{c}{\underline{\phm{aaa}Intrinsic\phm{aaa}}} &
           \multicolumn{2}{c}{\underline{\phm{aaa}Apparent\phm{aaa}}}\tm{a} & 
  S/N\tm{b} & \multicolumn{3}{c}{\underline{
   \phm{aaaaaaaaaaaaaa}$<N_a/N_{true}>$\phm{aaaaaaaaaaaaaa}}}\\
FWHM & Depth & FWHM & Depth & & 3 pixel rebin & 5 pixel rebin & 10
pixel rebin\\
 (km\,s$^{-1}$) & (\%) & (km\,s$^{-1}$) & (\%) & & (6\kms\ bins) &
(10\kms\ bins) & (20\kms\ bins) } 
\startdata
40.0 & 20 & 45.8 & 18 & 10 &  1.20$\pm$0.51 & 1.11$\pm$0.49 & 1.07$\pm$0.52 \\
     &    &      &    &  8 &  1.32$\pm$0.69 & 1.16$\pm$0.64 & 1.09$\pm$0.68 \\
     &    &      &    &  6 &  1.73$\pm$1.05 & 1.37$\pm$0.91 & 1.24$\pm$0.91 \\
     &    &      &    &  4 &  3.18$\pm$2.14 & 2.02$\pm$1.70 & 1.49$\pm$1.47 \\
40.0 & 50 & 47.8 & 45 & 10 &  1.07$\pm$0.20 & 1.04$\pm$0.19 & 1.01$\pm$0.20 \\
     &    &      &    &  8 &  1.13$\pm$0.27 & 1.05$\pm$0.24 & 1.01$\pm$0.24 \\
     &    &      &    &  6 &  1.29$\pm$0.44 & 1.13$\pm$0.38 & 1.04$\pm$0.35 \\
     &    &      &    &  4 &  1.92$\pm$0.77 & 1.49$\pm$0.70 & 1.20$\pm$0.56 \\
40.0 & 99 & 61.1 & 97 & 10 &  0.87$\pm$0.07 & 0.85$\pm$0.07 & 0.80$\pm$0.08 \\
     &    &      &    &  8 &  0.90$\pm$0.09 & 0.87$\pm$0.09 & 0.81$\pm$0.10 \\
     &    &      &    &  6 &  0.93$\pm$0.10 & 0.89$\pm$0.11 & 0.83$\pm$0.12 \\
     &    &      &    &  4 &  1.00$\pm$0.16 & 0.94$\pm$0.16 & 0.85$\pm$0.17 \\
\tableline
20.0 & 20 & 28.6 & 14 & 10 &  1.21$\pm$0.81 & 1.11$\pm$0.81 & 1.04$\pm$0.87 \\
     &    &      &    &  8 &  1.35$\pm$1.03 & 1.19$\pm$1.01 & 1.05$\pm$1.06 \\
     &    &      &    &  6 &  1.78$\pm$1.57 & 1.41$\pm$1.44 & 1.18$\pm$1.44 \\
     &    &      &    &  4 &  3.42$\pm$3.35 & 2.15$\pm$2.71 & 1.45$\pm$2.34 \\
20.0 & 50 & 29.3 & 37 & 10 &  1.03$\pm$0.30 & 0.99$\pm$0.31 & 0.94$\pm$0.30 \\
     &    &      &    &  8 &  1.11$\pm$0.42 & 1.04$\pm$0.41 & 0.97$\pm$0.40 \\
     &    &      &    &  6 &  1.27$\pm$0.62 & 1.12$\pm$0.56 & 0.99$\pm$0.52 \\
     &    &      &    &  4 &  2.06$\pm$1.34 & 1.56$\pm$1.18 & 1.18$\pm$0.92 \\
20.0 & 99 & 34.2 & 89 & 10 &  0.67$\pm$0.09 & 0.63$\pm$0.08 & 0.56$\pm$0.07 \\
     &    &      &    &  8 &  0.69$\pm$0.12 & 0.65$\pm$0.11 & 0.57$\pm$0.08 \\
     &    &      &    &  6 &  0.75$\pm$0.16 & 0.69$\pm$0.15 & 0.58$\pm$0.12 \\
     &    &      &    &  4 &  0.88$\pm$0.25 & 0.79$\pm$0.26 & 0.64$\pm$0.23 \\
\enddata
\tablecomments{The \fuse/LiF spectrograph has FWHM$_{ins}$=20\kms\ and
  pixel size 2.0\kms. The upper and lower halves of this table
  represent fullly resolved and marginally resolved lines, respectively.} 
\tn{a}{Instrumental broadening causes the apparent line depth to be less
  than the intrinsic line depth, and the apparent width to be broader
  than the intrinsic line width.}
\tn{b}{S/N is quoted per 20\kms\ resolution element.}
\end{deluxetable}

\begin{deluxetable}{lcccc ccc}
\tablewidth{0pt}
\tabcolsep=4pt
\tabletypesize{\scriptsize}
\tablecaption{Simulation Results - STIS/E140M}
\tablehead{\multicolumn{2}{c}{\underline{\phm{aaa}Intrinsic\phm{aaa}}} &
           \multicolumn{2}{c}{\underline{\phm{aaa}Apparent\phm{aaa}}}\tm{a} & 
  S/N\tm{b} & \multicolumn{3}{c}{\underline{
   \phm{aaaaaaaaaaaaaa}$<N_a/N_{true}>$\phm{aaaaaaaaaaaaaa}}}\\
FWHM & Depth & FWHM & Depth & & 1 pixel rebin & 2 pixel rebin & 3
pixel rebin\\
 (km\,s$^{-1}$) & (\%) & (km\,s$^{-1}$) & (\%) & & (3.2\kms\ bins) &
(6.4\kms\ bins) & (9.6\kms\ bins) } 
\startdata
13.6 & 20 & 15.6 & 18 & 10 &  1.11$\pm$0.52 & 1.04$\pm$0.54 & 1.03$\pm$0.53 \\
     &    &      &    &  8 &  1.16$\pm$0.70 & 1.04$\pm$0.70 & 1.02$\pm$0.71 \\
     &    &      &    &  6 &  1.48$\pm$1.00 & 1.28$\pm$0.97 & 1.21$\pm$0.94 \\
     &    &      &    &  4 &  2.11$\pm$1.81 & 1.46$\pm$1.52 & 1.29$\pm$1.43 \\
13.6 & 50 & 16.2 & 45 & 10 &  1.03$\pm$0.18 & 1.00$\pm$0.19 & 0.98$\pm$0.19 \\
     &    &      &    &  8 &  1.06$\pm$0.25 & 1.01$\pm$0.24 & 0.98$\pm$0.23 \\
     &    &      &    &  6 &  1.11$\pm$0.36 & 1.02$\pm$0.34 & 0.99$\pm$0.33 \\
     &    &      &    &  4 &  1.44$\pm$0.69 & 1.14$\pm$0.53 & 1.06$\pm$0.48 \\
13.6 & 99 & 20.8 & 97 & 10 &  0.87$\pm$0.08 & 0.82$\pm$0.08 & 0.75$\pm$0.07 \\
     &    &      &    &  8 &  0.89$\pm$0.09 & 0.84$\pm$0.09 & 0.76$\pm$0.09 \\
     &    &      &    &  6 &  0.91$\pm$0.11 & 0.85$\pm$0.12 & 0.78$\pm$0.13 \\
     &    &      &    &  4 &  0.96$\pm$0.16 & 0.87$\pm$0.17 & 0.81$\pm$0.18 \\
\tableline
 6.8 & 20 &  9.7 & 14 & 10 &  1.10$\pm$0.78 & 1.02$\pm$0.81 & 1.01$\pm$0.85 \\
     &    &      &    &  8 &  1.21$\pm$1.05 & 1.09$\pm$1.08 & 1.05$\pm$1.16 \\
     &    &      &    &  6 &  1.43$\pm$1.45 & 1.24$\pm$1.48 & 1.17$\pm$1.56 \\
     &    &      &    &  4 &  2.50$\pm$2.98 & 1.76$\pm$2.48 & 1.55$\pm$2.35 \\
 6.8 & 50 & 10.0 & 37 & 10 &  0.98$\pm$0.28 & 0.94$\pm$0.30 & 0.92$\pm$0.31 \\
     &    &      &    &  8 &  1.04$\pm$0.37 & 0.99$\pm$0.37 & 0.95$\pm$0.38 \\
     &    &      &    &  6 &  1.09$\pm$0.51 & 0.98$\pm$0.49 & 0.93$\pm$0.51 \\
     &    &      &    &  4 &  1.52$\pm$1.10 & 1.18$\pm$0.87 & 1.09$\pm$0.86 \\
 6.8 & 99 & 11.6 & 90 & 10 &  0.66$\pm$0.09 & 0.60$\pm$0.07 & 0.53$\pm$0.07 \\
     &    &      &    &  8 &  0.67$\pm$0.12 & 0.60$\pm$0.10 & 0.52$\pm$0.09 \\
     &    &      &    &  6 &  0.72$\pm$0.16 & 0.62$\pm$0.14 & 0.54$\pm$0.12 \\
     &    &      &    &  4 &  0.84$\pm$0.26 & 0.69$\pm$0.25 & 0.58$\pm$0.21 \\
 \enddata
\tablecomments{STIS/E140M has FWHM$_{ins}$=6.8\kms\ and pixel size
  3.2\kms. The upper and lower halves of this table
  represent fullly resolved and marginally resolved lines, respectively.}
\tn{a}{Instrumental broadening causes the apparent line depth to be less
  than the intrinsic line depth, and the apparent width to be broader
  than the intrinsic line width.}
\tn{b}{S/N is quoted per 6.8\kms\ resolution element.}
\end{deluxetable}

\begin{figure}
\epsscale{0.9}
\plotone{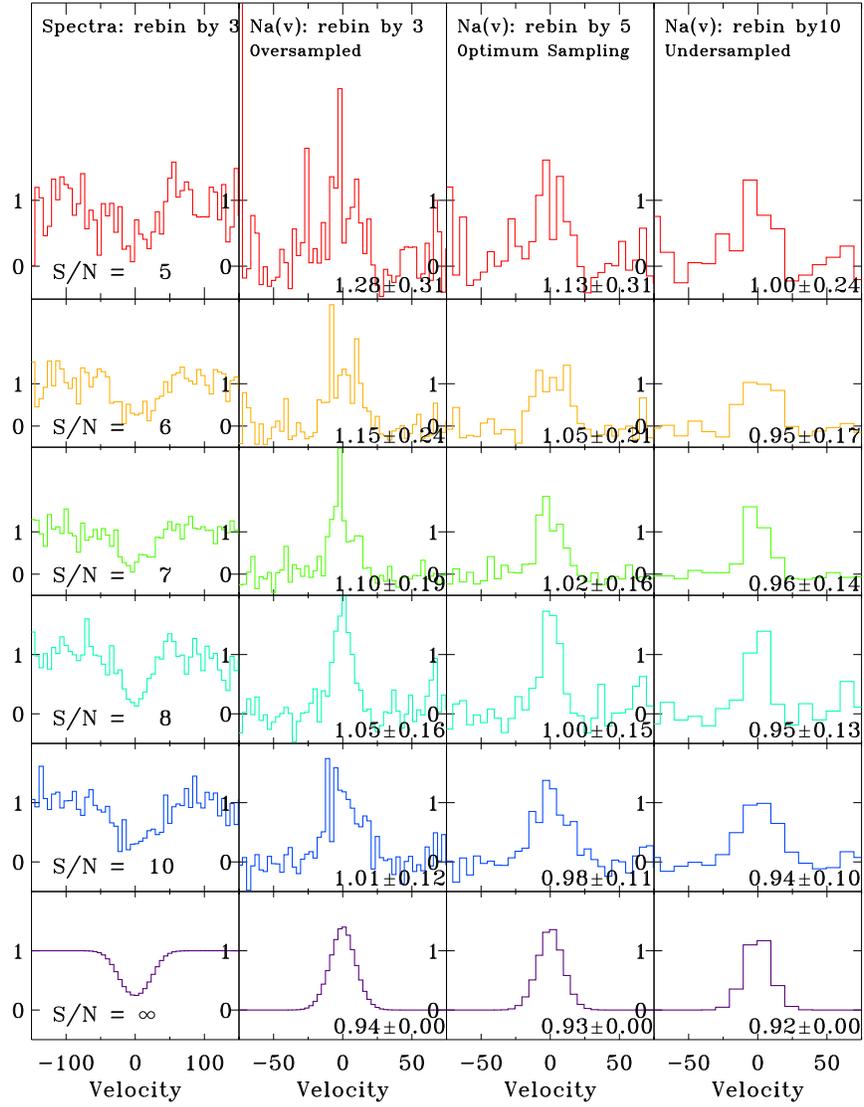}
\caption{Simulated spectra illustrating the effect by which $N_a$ is
  overestimated when the AOD technique is applied to noisy data, for
  resolved absorption lines observed with \fuse\ 
  with an intrinsic central depth of 80\%. 
  The left column shows the simulated spectra with various levels of
  Poisson noise added. The next three columns show the effect of  
  translating the noise into apparent column density space, after different
  levels of rebinning. The $N_a(v)$ profiles have been arbitrarily
  normalized for convenience. The numbers written in the corner of each
  panel represent the mean column density overestimation factor
  $<\!N_a\!>/N_{true}$ after 500 runs, together with its dispersion, where
  $N_a=\int N_a(v)\mathrm{d}v$ and $N_{true}$ is the true column
  density in the absorber. S/N is quoted per resolution element
  (corresponding in this case to the S/N at 10 pixel rebinning).}
\end{figure}

\begin{figure}
\epsscale{0.9}
\plotone{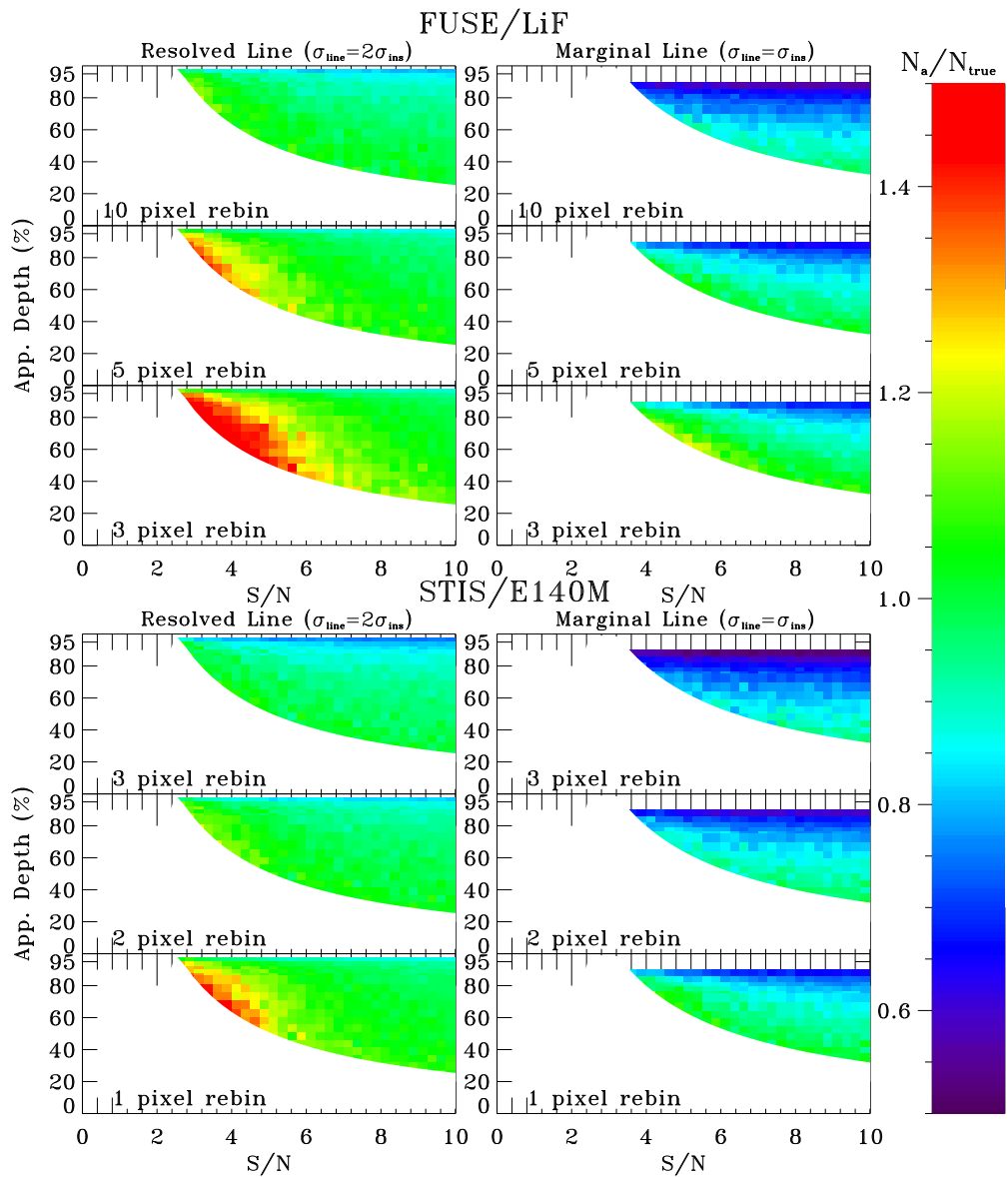}
\caption{Color-coded plots showing the column density overestimation
  factor $N_a/N_{true}$ as a function of signal-to-noise per
  resolution element, apparent
  line depth, and rebinning, for absorption lines measured using the AOD
  method in four cases: \fuse\ resolved lines (top-left), \fuse\
  marginally resolved lines (top-right), STIS resolved lines
  (bottom-left), and STIS marginally resolved lines
  (bottom-right). Only lines with $>3\sigma$ equivalent width detections 
  are included. Green regions show cases where the AOD method returns
  the column density correct to within 20\%; blue and purple
  regions show where the AOD method will
  underestimate the true column density (unresolved lines,
  particularly at high depth); yellow and red regions show cases where
  AOD measurements 
  lines will overestimate the column density at (low S/N, low depth).}
\end{figure}

\begin{figure}
\epsscale{0.5}
\plotone{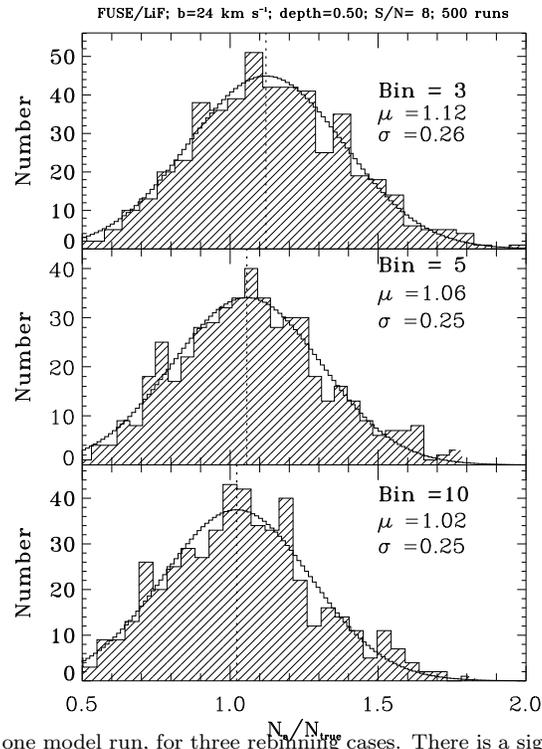}
\caption{Histogram of $N_a/N_{true}$ for one model run, for three
  rebinning cases. There is a significant dispersion in overestimation
  factors for a given case. As the rebinning increases, the most likely value
  of $N_a/N_{true}$ approaches unity}
\end{figure}

\end{document}